\documentclass{article}

\usepackage{arxiv}
\usepackage{amssymb}
\usepackage{latexsym}
\usepackage{amssymb}
\usepackage{amsmath}
\usepackage{url}
\usepackage{xcolor}

\usepackage{subfigure}
\usepackage{multirow}
\usepackage{caption}
\usepackage[utf8]{inputenc} % allow utf-8 input
\usepackage[T1]{fontenc}    % use 8-bit T1 fonts
\usepackage{hyperref}       % hyperlinks
\usepackage{url}            % simple URL typesetting
\usepackage{booktabs}       % professional-quality tables
\usepackage{amsfonts}       % blackboard math symbols
\usepackage{nicefrac}       % compact symbols for 1/2, etc.
\usepackage{microtype}      % microtypography
\usepackage{cleveref}       % smart cross-referencing
\usepackage{lipsum}         % Can be removed after putting your text content
\usepackage{graphicx}
\usepackage{natbib}
\usepackage{doi}
\usepackage{booktabs}
\usepackage{multirow}

\title{Image Restoration using Feature-guidance}

\date{}

\author{ \hspace{1mm} Maitreya Suin
	\And
	\hspace{1mm} Kuldeep Purohit
	\And
	\hspace{1mm} A. N. Rajagopalan
}

\renewcommand{\headeright}{}
\renewcommand{\undertitle}{}
\begin{document}
\maketitle

\begin{abstract}
Image restoration is the task of recovering a clean image from a degraded version. In most cases, the degradation is spatially varying, and it requires the restoration network to both localize and restore the affected regions. In this paper, we present a new approach suitable for handling the image-specific and spatially-varying nature of degradation in images affected by practically occurring artifacts such as blur,  rain-streaks. We decompose the restoration task into two stages of degradation localization and degraded region-guided restoration, unlike existing methods which directly learn a mapping between the degraded and clean images. Our premise is to use the auxiliary task of degradation mask prediction to guide the restoration process. We demonstrate that the model trained for this auxiliary task contains vital region knowledge, which can be exploited to guide the restoration network's training using attentive knowledge distillation technique. Further, we propose mask-guided convolution and global context aggregation module that focuses solely on restoring the degraded regions. The proposed approach's effectiveness is demonstrated by achieving significant improvement over strong baselines.
\end{abstract}

\section{Introduction}
Image restoration is a task of learning a mapping function that transforms input degraded images into enhanced images devoid of the artifacts and distortions. Due to the complexity of scene contents and weather conditions, image degradation is often inevitable. Such degradations impact not only human visibility but also degrade many computer vision applications, such as autonomous driving, drone flying, and surveillance systems, etc. As in many other computer vision tasks, the employment of deep convolutional networks has made significant progress. Given a large number of original and restored image pairs, the task can be solved by image-to-image translation methods, which have made considerable progress. In this study, aiming at further improvements, we pursue a different design that explores the importance of degraded-region detection and show that this approach is equally beneficial for various tasks of image restoration.

Existing CNN-based methods for image restoration employ a single network while focusing on the effectiveness of different baseline architectures \cite{jiao2017formresnet,zhang2019deep}, enlarging the receptive field \cite{liu2018non}, using multi-scale information \cite{fan2019scale}, feature exploitation etc. However, the fact that the spatial scales and orientations of the degradations may vary across an image is mostly overlooked by existing methods. Intuitively, image restoration can be explained as a combination of two tasks - detection and then the restoration of affected parts of an image. In this work we address the task of removing complex spatially varying image degradations. Such degradations usually arise due to the medium between the camera and the scene and the dynamics of the scene with respect to camera. Bad weather causes significant image quality and visibility degradation in form of rain.These tasks are highly spatially varying due to non uniform depth variations, rain streaks' directions and locations (rain-streak). Directly regressing from degraded to clean image becomes a challenging task, and training a single network with this objective limits the restoration quality. We claim that it is equally necessary to accurately localize the regions which are affected and having prior knowledge of the affected areas significantly helps a restoration network. Few recent approaches also have explored the utility of auxiliary information like edge \cite{nazeri2019edgeconnect}, depth-map \cite{li2020dynamic}, etc., which are tailored for a particular task. For constructing a general approach that will be equally effective for any of the spatially varying restoration tasks, instead of using any particular guidance like object boundaries, segmentation, etc., we propose to localize the affected regions first and then use this region knowledge to guide the restoration process effectively. We demonstrate the efficacy of our approach for removing these atmospheric degradations and claim that it can be equally effective for any restoration task with highly spatially-varying nature, and we choose dynamic scene deblurring as an additional application, where depth variations and moving objects results in spatially varying degradation.

Following \cite{wang2019spatial,qian2018attentive}, we take the difference between the ground-truth and the input image and threshold it to generate the ground-truth degradation masks. Intuitively these masks denote the regions with significant degradation, which are difficult to restore. In the first step, we pre-train a mask prediction module with the same structure as the restoration network but with much fewer layers for a simpler binary classification task. Next, we focus on extracting the knowledge from the mask-prediction network and improve the original restoration task.

We propose to use attentive-distillation technique to extract relevant intermediate feature-level information from the mask prediction network and try to transfer this knowledge to the restoration network. Intuitively, it can be related to a type of transfer learning problem where the target domain is identical to the source domain, but the target task is different from the source task \cite{pan2009survey}. We show that features extracted by the mask prediction network can act as better localization cues and can be exploited as additional supervision to the restoration-encoder (Sec. \ref{sec:knowledge_transfer}). The intuition behind this design choice is that the mask-encoder should excel at extracting features with rich information about the degraded regions, and the restoration-encoder can learn its behavior to obtain more useful features with better region information. We further allow the restoration encoder to adaptively calibrate its attention towards only the relevant and most useful features of the mask-encoder.

We also demonstrate that the predicted mask can be exploited to improve the restoration performance further while requiring slightly more parameters for the mask prediction network at inference time. Apart from implicit knowledge transfer from the mask prediction network, the mask itself can be utilized in the decoder to help restoration. We show the efficacy of mask-guided gated convolutional operation, which allows the layer to put more attention on the degraded regions to process it with greater detail. 

To summarize, our contributions are as follows: (1) We propose to explicitly disentangle the general image restoration task into region localization and region-guided restoration. To the best of our knowledge, our work is among the first of a few studies that show the efficacy of region knowledge for general restoration problem. (2) We propose the use of an attentive knowledge distillation technique to transfer the region-knowledge to the restoration network, which improves the performance without requiring any extra computational overhead at inference. (3) Further, we show the efficacy of mask-guided modules in the decoder to guide the restoration process with the predicted mask explicitly. (4) Our experiments demonstrate the effectiveness of the proposed approach for multiple restoration tasks. We claim these techniques can be incorporated for any general spatially varying image restoration problem to improve the performance significantly. 
\section{Related Works}

\textbf{Rain-streak removal}
Conventional image deraining methods adopt a model-driven methodology utilizing physical properties of rain and prior knowledge of background scenes into an optimization problem. Early approaches proposed use of pre-defined filters  \cite{ding2016single}, priors \cite{zhu2017joint}, layer separation model \cite{li2016rain}, or screen blend model \cite{luo2015removing}. CNN-based methods significantly advanced the state-of-the-art in deraining.\cite{fu2017removing} proposed a synthetic large-scale rain-streak dataset and used it to learn an end-to-end negative residual mapping using a 3-layer CNN. \cite{yang2017deep} constructed a more diversified dataset and deeper CNN architecture which takes advantage of larger receptive field using dilation filters, improving the deraining results by jointly detecting and removing rain-streaks.

\textbf{Motion blur removal}
Motion deblurring is a challenging problem in computer vision due to its ill-posed nature. Major efforts have gone into designing priors that are apt for recovering the underlying undistorted image and the camera trajectory \cite{lai2016comparative}. However, these methods preclude commonly occurring real-world blur arises from various sources including moving objects, camera shake and depth variations, causing different pixels to acquire different motion trajectories. A significant number of works have been proposed (\cite{paramanand2011depth,nimisha2018unsupervised,rao2014harnessing,nimisha2018generating,vasu2017local,paramanand2014shape,vijay2013non}) where various traditional approaches were adopted for deblurring. Recent works (\cite{purohit2019bringing,purohit2020region,mohan2021deep,mohan2019unconstrained,vasu2018non}) based on deep convolutional neural networks (CNN) have studied the benefits of replacing the image formation model with a parametric model that can be trained to emulate the non-linear relationship between blurred-sharp image pairs. 
Similar approaches can be found for image super-resolution (\cite{suresh2007robust,vasu2018analyzing,bhavsar2010resolution,bhavsar2012range,rajagopalan2005background,rajagopalan2003motion,nimisha2021blind,purohit2020mixed,punnappurath2017multi,punnappurath2015rolling}) and rolling shutter deblurring (\cite{rengarajan2017unrolling,rengarajan2016bows,kandula2020deep,vasu2018occlusion,vasu2017camera,mohan2017going,rengarajan2016image,pichaikuppan2014change}). Locally linear blur kernel assumption is explored in \cite{sun2015learning,gong2017motion} with limited success in general dynamic scenes. Use of fully convolutional CNNs to directly estimate the latent sharp image was proposed in~\cite{nah2017deep} and adopted by recent works.  \cite{nah2017deep,tao2018scale,gao2019dynamic} proposed encoder-decoder residual networks to aggregate features in a coarse-to-fine manner, while showing benefits of selective parameter/feature sharing and/or recurrent layers. \cite{zhang2018dynamic} explored a design composed of multiple CNNs and RNN. Recently, \cite{zhang2019deep} proposed a multi-patch hierarchical network and stacked its copies along depth to achieve state-of-the-art performance. 

\section{Method}
One of the most widely employed network architectures in image-to-image tasks is the encoder-decoder architecture \cite{ronneberger2015u}. It has showed its efficacy in image inpainting \cite{liu2018image}, semantic image segmentation \cite{isola2017image}, image deblurring \cite{tao2018scale}, etc. In this paper, to make our approach as general as possible, we select a standard encoder-decoder model as the baseline for all the restoration tasks addressed. A detailed layerwise description of the network is given in supplementary material. Similarly, we deploy a very lightweight version of the encoder-decoder model (with similar level and structure) as the mask prediction network. This is because the task of binary classification is much simpler than the intensity regression task. We first train the mask prediction module using binary cross-entropy loss to predict the severely degraded regions. On the second step, we train a restoration network that actually carries out the restoration process utilizing the already available knowledge of the degraded regions. We also use a pixel-shuffling layer at the beginning that transforms the image pixels to channel-space using pixel-shuffling by a factor of 2. This allows subsequent computationally intensive operations to be performed at lower spatial resolution. 
\subsection{\textbf{Knowledge-transfer}}
\label{sec:knowledge_transfer}
Information transferability in different layers was explored by  \cite{yosinski2014transferable}, e.g., the first layers learn general features like shape or structure, the middle layers learn high-level semantic features, and the last layers learn the features that are very specific to a particular task. We mainly focus on the encoders to convey knowledge about the regions that need to be restored. If the outputs from the $l^{th}$ layer of the mask encoder and restoration encoder are $x_l^*$ and $x_l$, then regularizing term can be defined as:
\begin{equation}
	\label{basic_distil}
	R(\theta,x)^{l,l} = ||x_l - x_l^* ||^2_2
\end{equation}
where $x$ is the input, and $\theta$ is the parameters of the restoration network. We regularize the ``Behavior,'' i.e., feature maps rather than model parameters. We want the guided layer in the restoration encoder to extract the relevant information from the hint layer of the mask-encoder using knowledge distillation. We use the down-sampling layers of the network as the breakpoints.
\begin{figure*}[!t]
	\centering
	\includegraphics[width = 0.8\textwidth]{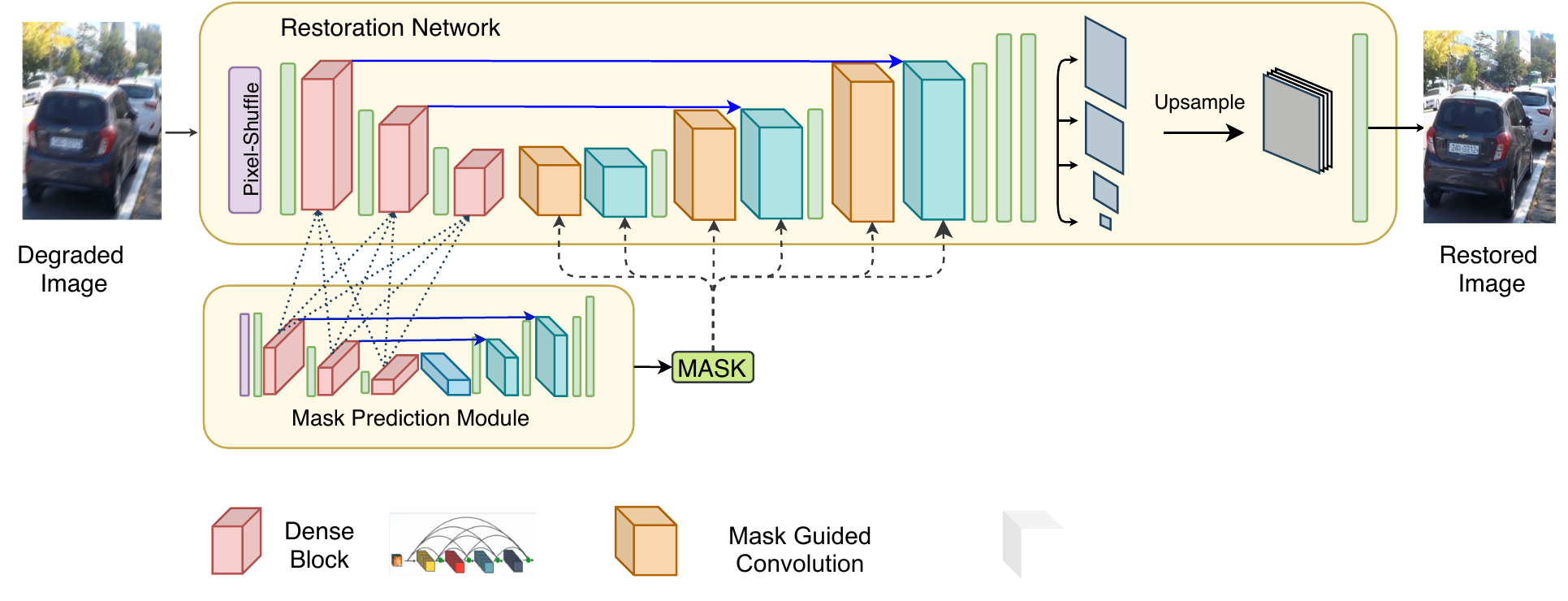}
	\caption{Proposed network and its components.}
	\label{fig:att_distil}
	\vspace{-0.5cm}    
\end{figure*}
%\vspace{1cm}

We also use meta-networks \cite{jang2019learning} to decide which feature maps (channels) of the mask model are useful and relevant for the particular restoration task and which mask layers should be transferred to which restoration layers (Fig. \ref{fig:att_distil}). If there are $m$ encoder level/breakpoints, for each of the $m \times m$ pair, we introduce transfer importance predictor, which enforces different penalties for each channel according to their utility on the target task. For any particular pair, Eq. \ref{basic_distil} can be modified as
\begin{equation}
	R(\theta,x,\rho^{p,q})^{p,q} = \sum_{c\in C} \rho^{p,q}_c(x_q^*) ||(x_p - x_q^*)_c ||^2_2
\end{equation}
where $p,q \in \{ 1,m\}$ and $\rho^{p,q}_c : \mathbb{R}^{C \times H \times W} \rightarrow \mathbb{R}^{C}$ is the non-negative weight of channel c with $\sum_{c\in C} \rho^{p,q}_c = 1$. For any tensor $z$, the term $z_c$ denotes the $c^{th}$ slice of the tensor. Similarly we denote the importance of each pair as $\alpha^{p,q} : \mathbb{R}^{C \times H \times W} \rightarrow \mathbb{R}^{1} \geq 0$. The combined transfer loss given the weights of channels $\rho$ and weights of matching pairs $\alpha$ is
\begin{equation}
	\label{eqn:distil_final}
	R = \sum_{(p,q)\in (m,m)} \alpha^{p,q}(x^*_q) R(\theta,x,\rho^{l,l})^{p,q}
\end{equation}
Note that we suitably apply spatial interpolation/$1\times 1$ convolution to match the feature pair's dimension.
\subsection{\textbf{Mask-guided Convolution}}
Feature-wise masking has been explored extensively in many tasks. \cite{hu2018squeeze}  re-calibrate feature responses by explicitly multiplying each channel with a learned sigmoidal mask. Recently in image inpainting, few methods advocate using masked convolution operation to reduce the impact of invalid pixels or holes. \cite{liu2018image} categorizes all input locations as either invalid or valid and multiplies a zero-or-one mask to inputs throughout all layers. For general image restoration task, we demonstrate the efficacy of gated convolution technique where we use the predicted mask to focus on restoring the degraded regions. Given the input feature map, $x$ and the predicted-mask $M$, the output can be calculated as 
\begin{equation}
	f' = (W^Tf) \odot M
\end{equation}
where $W$ is the weight of the convolutional layer.
\section{Experiments}
\subsection{Training description:} The restoration network minimizes $l_1$ reconstruction loss between the network output and the GT clean image along with attentive distillation loss (Eq. \ref{eqn:distil_final}). The degradation-mask estimator is trained using binary cross entropy loss with respect to the GT binary mask. Each training batch contains randomly cropped RGB patches of size $256\times256$ from degraded images and we randomly flip them horizontally or vertically as the inputs. The batch-size was 8 for rain-streak removal and 16 for deblurring. Both stages use Adam optimizer with initial leaning rate $10^{-4}$, halved after every $50$ epochs. We use PyTorch library and Titan Xp GPU.

\subsection{Datasets}
\noindent \textbf{Rain-streak:} 
We utilize a challenging benchmark datasets in our experiments on rain streak removal viz. Rain100H \cite{yang2017deep}. Rain100H dataset contains of $1800$ labeled images for training and $100$ images for testing. Following the training and testing dataset split described in \cite{yang2017deep}, our model is evaluated quantitatively using PSNR and SSIM scores on the luminance channel on Rain100H dataset.

\noindent \textbf{Motion blur:} 
We follow the configuration of~\cite{zhang2019deep,kupyn2019deblurgan,tao2018scale,kupyn2017deblurgan,nah2017deep}, which train on 2103 images from the GoPro dataset~\cite{nah2017deep}. Also for testing, we use: GoPro~\cite{nah2017deep} (1103 HD images).

\subsection{Rain-streak Removal}
\label{sec:rain}
\begin{figure*}[htb]
	\centering
	\includegraphics[width = 0.95\textwidth]{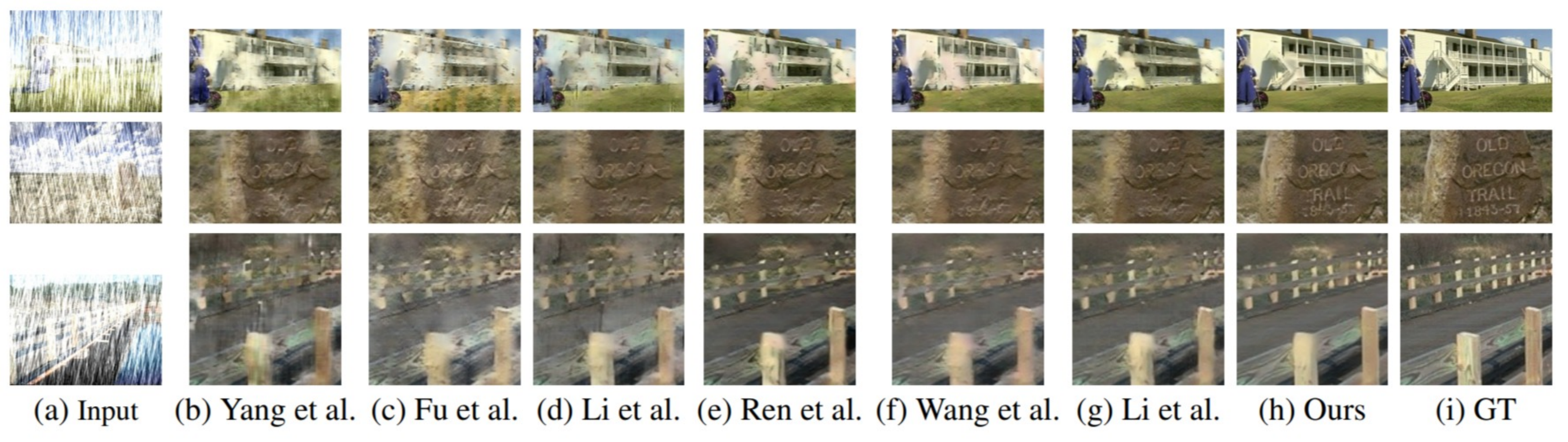}
	\caption{Qualitative comparison of zoomed-in results on synthetic rainy images from the Rain100H test-set.}
	\label{fig:results_heavy}
	%\vspace{-9mm}
\end{figure*}

Various existing methods are included in comparisons, including representative traditional methods including DSC \cite{luo2015removing} and GMM \cite{li2016rain}, and several state-of-the-art deep CNN-based models i.e., DDN \cite{fu2017removing}, JORDER \cite{yang2017deep}, DDN \cite{fu2017removing}, JBO\cite{zhu2017joint}, DID-MDN\cite{zhang2018density}, RESCAN \cite{li2018recurrent}, PreNet \cite{ren2019progressive}, and SPANet \cite{wang2019spatial}. Derained image results provided by the respective authors were used for Rain100H dataset. 

As can be inferred from Table \ref{table:syn_rain100h}, our network achieves significant PSNR gains over all the competing methods. Learning-based deraining methods perform better than traditional techniques. Representative results from selected test images from Rain100H dataset are provided in Fig. \ref{fig:results_heavy}. It can be seen that the visual quality of our results is significantly higher than that of state-of-the-art methods which contain visible rain streaks or missing textured regions.

\begin{table*}[t]%[t]
% \vspace{-2mm}
\centering
\resizebox{\textwidth}{!}{%
\begin{tabular}{@{}ccccccccccc@{}}
\toprule
Method &DSC  & GMM  & Clear & DDN  & JORDER   & DID-MDN & SPANet & NLEDN & PreNet  &
\textbf{Ours}  \\ \midrule
PSNR & 17.55 & 15.96 & 22.07 & 21.91 & 26.54 & 26.35 & 29.71 & 30.38 & 29.46 & \textbf{31.93} \\ \midrule
SSIM & 0.538 & 0.418 & 0.842 & 0.887 & 0.835 & 0.829 & 0.8522 & 0.894 & 0.893 & \textbf{0.924} \\ \bottomrule
\end{tabular}%
}
\caption{Quantitative evaluation and comparisons on the Rain100H benchmark.}
\label{table:syn_rain100h}
\end{table*}

\subsection{Motion Blur Removal from Dynamic Scenes}
\label{sec:motion-blur}

\begin{table*}[htbp]
\centering
\caption{Performance comparisons with existing algorithms on 1103 images from the deblurring benchmark GoPro \cite{nah2017deep}.\label{TableGopro}}
\resizebox{\textwidth}{!}{
\begin{tabular}{c|c|c|c|c|c|c|c|c|c|c|c}
\hline
Method &  Whyte et al. & Hyun Kim et al. & Gong et al. & Nah et al. & Kupyn et al. & Tao et al. & Zhang et al. & Gao et al. & Zhang et al.& Kupyn et al. & Ours\\
\hline
PSNR (dB)  & 24.6 & 23.64 &  26.4  & 29.08 & 28.7 & 30.26 & 29.19 &30.90 & 31.20 &29.55 & \textbf{32.10} \\
SSIM  & 0.846 & 0.824 & 0.863 & 0.914 & 0.858 & 0.934 & 0.931 &0.935 &0.940 &0.934 & \textbf{0.956}\\
\hline
\end{tabular}
}
\end{table*}

%\subsection{Results}
We futher validate our approach for general dynamic scene deblurring. Due to the complexity of the blur present in such images, conventional image formation model based deblurring approaches struggle to perform well. We provide extensive comparisons with \cite{whyte2012non} (representative traditional non-uniform deblurring model) and state-of-the-art learning-based methods, namely MS-CNN\cite{nah2017deep}, DeblurGAN\cite{kupyn2017deblurgan}, DeblurGAN-v2\cite{kupyn2019deblurgan}, SRN\cite{tao2018scale}, and Stack(4)-DMPHN\cite{zhang2019deep}. Official implementation from the authors were used with default parameters.

%\textbf{Quantitative Evaluation:}
The average PSNR and SSIM scores on GoPro test set are listed in Table \ref{TableGopro}. We compare against all existing models trained on GoPro train-set for fair comparisons. 

\begin{figure*}[htb] \label{fig:visual_deblur}
	\centering
	\includegraphics[width = 0.95\textwidth]{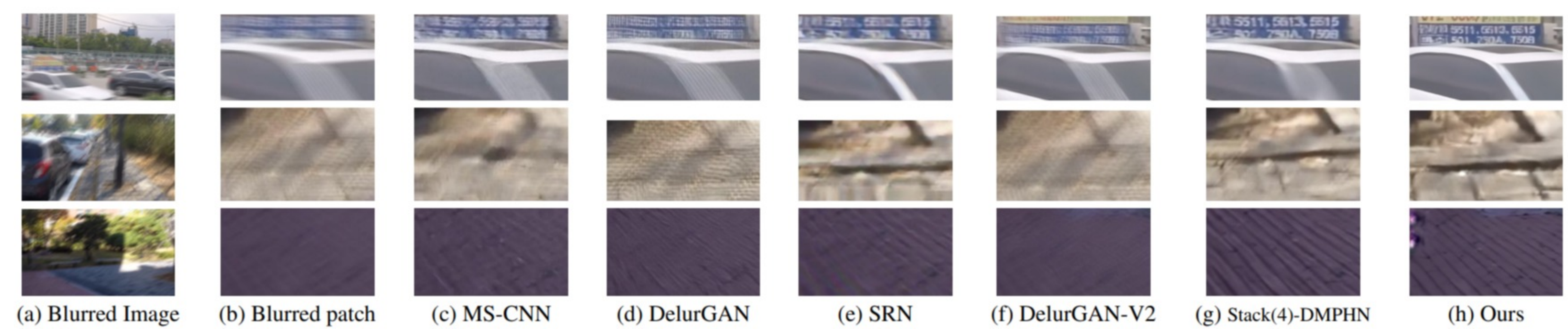}
	%\vspace{-0em}
	\caption{Visual comparisons of deblurring results on images from the GoPro test set~\cite{nah2017deep}. Key blurred patches are shown in (b), while zoomed-in patches from the deblurred results are shown in (c)-(h).}% (best viewed in high resolution).}
\label{fig:dynamic}
	%\vspace{-0em}
\end{figure*}

%\textbf{Qualitative Evaluation:} 
Visual comparisons on images containing dynamic and 3D scenes are shown in Figs.~\ref{fig:dynamic}. We observe that the results of prior works suffer from incomplete deblurring or artifacts. In contrast, our network demonstrates large dynamic blur handling capability while preserving sharpness. Scene details in the regions containing text, object boundaries, and textures are more faithfully restored, making them recognizable.

\section{Conclusion}
We addressed the task of removing degradations from an image and show
that our approach can generalize well to any image restoration
task that is spatially varying in nature. We take an off-the-shelf encoder-decoder
architecture as our strong backbone. We model the restoration
task as a combination of degraded-region segmentation and
region guided restoration. We propose a distillation technique
for image restoration where we leverage the knowledge of a
degradation mask prediction module as extra guidance. We
show that this training strategy achieves superior results over
the baseline for all of the tasks addressed without requiring
any extra parameters at inference. Refined and complete version of this work appeared in IEEE-JSTSP.

\bibliographystyle{unsrtnat}
\bibliography{references}  %%% Uncomment this line and comment out the ``thebibliography'' section below to use the external .bib file (using bibtex) .

\begin{thebibliography}{64}
\providecommand{\natexlab}[1]{#1}
\providecommand{\url}[1]{\texttt{#1}}
\expandafter\ifx\csname urlstyle\endcsname\relax
  \providecommand{\doi}[1]{doi: #1}\else
  \providecommand{\doi}{doi: \begingroup \urlstyle{rm}\Url}\fi

\bibitem[Jiao et~al.(2017)Jiao, Tu, He, and Lau]{jiao2017formresnet}
Jianbo Jiao, Wei-Chih Tu, Shengfeng He, and Rynson~WH Lau.
\newblock Formresnet: Formatted residual learning for image restoration.
\newblock In \emph{Proceedings of the IEEE Conference on Computer Vision and
  Pattern Recognition Workshops}, pages 38--46, 2017.

\bibitem[Zhang et~al.(2019)Zhang, Dai, Li, and Koniusz]{zhang2019deep}
Hongguang Zhang, Yuchao Dai, Hongdong Li, and Piotr Koniusz.
\newblock Deep stacked hierarchical multi-patch network for image deblurring.
\newblock In \emph{Proceedings of the IEEE Conference on Computer Vision and
  Pattern Recognition}, pages 5978--5986, 2019.

\bibitem[Liu et~al.(2018{\natexlab{a}})Liu, Wen, Fan, Loy, and
  Huang]{liu2018non}
Ding Liu, Bihan Wen, Yuchen Fan, Chen~Change Loy, and Thomas~S Huang.
\newblock Non-local recurrent network for image restoration.
\newblock In \emph{Advances in Neural Information Processing Systems}, pages
  1673--1682, 2018{\natexlab{a}}.

\bibitem[Fan et~al.(2019)Fan, Yu, Liu, and Huang]{fan2019scale}
Yuchen Fan, Jiahui Yu, Ding Liu, and Thomas~S Huang.
\newblock Scale-wise convolution for image restoration.
\newblock \emph{arXiv preprint arXiv:1912.09028}, 2019.

\bibitem[Nazeri et~al.(2019)Nazeri, Ng, Joseph, Qureshi, and
  Ebrahimi]{nazeri2019edgeconnect}
Kamyar Nazeri, Eric Ng, Tony Joseph, Faisal Qureshi, and Mehran Ebrahimi.
\newblock Edgeconnect: Structure guided image inpainting using edge prediction.
\newblock In \emph{Proceedings of the IEEE International Conference on Computer
  Vision Workshops}, pages 0--0, 2019.

\bibitem[Li et~al.(2020)Li, Pan, Lai, Gao, Sang, and Yang]{li2020dynamic}
Lerenhan Li, Jinshan Pan, Wei-Sheng Lai, Changxin Gao, Nong Sang, and
  Ming-Hsuan Yang.
\newblock Dynamic scene deblurring by depth guided model.
\newblock \emph{IEEE Transactions on Image Processing}, 29:\penalty0
  5273--5288, 2020.

\bibitem[Wang et~al.(2019)Wang, Yang, Xu, Chen, Zhang, and
  Lau]{wang2019spatial}
Tianyu Wang, Xin Yang, Ke~Xu, Shaozhe Chen, Qiang Zhang, and Rynson~WH Lau.
\newblock Spatial attentive single-image deraining with a high quality real
  rain dataset.
\newblock In \emph{Proceedings of the IEEE Conference on Computer Vision and
  Pattern Recognition}, pages 12270--12279, 2019.

\bibitem[Qian et~al.(2018)Qian, Tan, Yang, Su, and Liu]{qian2018attentive}
Rui Qian, Robby~T Tan, Wenhan Yang, Jiajun Su, and Jiaying Liu.
\newblock Attentive generative adversarial network for raindrop removal from a
  single image.
\newblock In \emph{Proceedings of the IEEE Conference on Computer Vision and
  Pattern Recognition}, pages 2482--2491, 2018.

\bibitem[Pan and Yang(2009)]{pan2009survey}
Sinno~Jialin Pan and Qiang Yang.
\newblock A survey on transfer learning.
\newblock \emph{IEEE Transactions on knowledge and data engineering},
  22\penalty0 (10):\penalty0 1345--1359, 2009.

\bibitem[Ding et~al.(2016)Ding, Chen, Zheng, Huang, and Zeng]{ding2016single}
Xinghao Ding, Liqin Chen, Xianhui Zheng, Yue Huang, and Delu Zeng.
\newblock Single image rain and snow removal via guided l0 smoothing filter.
\newblock \emph{Multimedia Tools and Applications}, 75\penalty0 (5):\penalty0
  2697--2712, 2016.

\bibitem[Zhu et~al.(2017)Zhu, Fu, Lischinski, and Heng]{zhu2017joint}
Lei Zhu, Chi-Wing Fu, Dani Lischinski, and Pheng-Ann Heng.
\newblock Joint bi-layer optimization for single-image rain streak removal.
\newblock In \emph{Proceedings of the IEEE international conference on computer
  vision}, pages 2526--2534, 2017.

\bibitem[Li et~al.(2016)Li, Tan, Guo, Lu, and Brown]{li2016rain}
Yu~Li, Robby~T Tan, Xiaojie Guo, Jiangbo Lu, and Michael~S Brown.
\newblock Rain streak removal using layer priors.
\newblock In \emph{Proceedings of the IEEE conference on computer vision and
  pattern recognition}, pages 2736--2744, 2016.

\bibitem[Luo et~al.(2015)Luo, Xu, and Ji]{luo2015removing}
Yu~Luo, Yong Xu, and Hui Ji.
\newblock Removing rain from a single image via discriminative sparse coding.
\newblock In \emph{Proceedings of the IEEE International Conference on Computer
  Vision}, pages 3397--3405, 2015.

\bibitem[Fu et~al.(2017)Fu, Huang, Zeng, Huang, Ding, and
  Paisley]{fu2017removing}
Xueyang Fu, Jiabin Huang, Delu Zeng, Yue Huang, Xinghao Ding, and John Paisley.
\newblock Removing rain from single images via a deep detail network.
\newblock In \emph{Proceedings of the IEEE Conference on Computer Vision and
  Pattern Recognition}, pages 3855--3863, 2017.

\bibitem[Yang et~al.(2017)Yang, Tan, Feng, Liu, Guo, and Yan]{yang2017deep}
Wenhan Yang, Robby~T Tan, Jiashi Feng, Jiaying Liu, Zongming Guo, and Shuicheng
  Yan.
\newblock Deep joint rain detection and removal from a single image.
\newblock In \emph{Proceedings of the IEEE Conference on Computer Vision and
  Pattern Recognition}, pages 1357--1366, 2017.

\bibitem[Lai et~al.(2016)Lai, Huang, Hu, Ahuja, and Yang]{lai2016comparative}
Wei-Sheng Lai, Jia-Bin Huang, Zhe Hu, Narendra Ahuja, and Ming-Hsuan Yang.
\newblock A comparative study for single image blind deblurring.
\newblock In \emph{Proceedings of the IEEE Conference on Computer Vision and
  Pattern Recognition}, pages 1701--1709, 2016.

\bibitem[Paramanand and Rajagopalan(2011)]{paramanand2011depth}
Chandramouli Paramanand and Ambasamudram~N Rajagopalan.
\newblock Depth from motion and optical blur with an unscented kalman filter.
\newblock \emph{IEEE Transactions on Image Processing}, 21\penalty0
  (5):\penalty0 2798--2811, 2011.

\bibitem[Nimisha et~al.(2018{\natexlab{a}})Nimisha, Sunil, and
  Rajagopalan]{nimisha2018unsupervised}
Thekke~Madam Nimisha, Kumar Sunil, and AN~Rajagopalan.
\newblock Unsupervised class-specific deblurring.
\newblock In \emph{Proceedings of the European Conference on Computer Vision
  (ECCV)}, pages 353--369, 2018{\natexlab{a}}.

\bibitem[Rao et~al.(2014)Rao, Rajagopalan, and Seetharaman]{rao2014harnessing}
Makkena~Purnachandra Rao, AN~Rajagopalan, and Guna Seetharaman.
\newblock Harnessing motion blur to unveil splicing.
\newblock \emph{IEEE transactions on information forensics and security},
  9\penalty0 (4):\penalty0 583--595, 2014.

\bibitem[Nimisha et~al.(2018{\natexlab{b}})Nimisha, Rajagopalan, and
  Aravind]{nimisha2018generating}
TM~Nimisha, AN~Rajagopalan, and Rangarajan Aravind.
\newblock Generating high quality pan-shots from motion blurred videos.
\newblock \emph{Computer Vision and Image Understanding}, 171:\penalty0 20--33,
  2018{\natexlab{b}}.

\bibitem[Vasu and Rajagopalan(2017)]{vasu2017local}
Subeesh Vasu and AN~Rajagopalan.
\newblock From local to global: Edge profiles to camera motion in blurred
  images.
\newblock In \emph{Proceedings of the IEEE Conference on Computer Vision and
  Pattern Recognition}, pages 4447--4456, 2017.

\bibitem[Paramanand and Rajagopalan(2014)]{paramanand2014shape}
Chandramouli Paramanand and AN~Rajagopalan.
\newblock Shape from sharp and motion-blurred image pair.
\newblock \emph{International journal of computer vision}, 107\penalty0
  (3):\penalty0 272--292, 2014.

\bibitem[Vijay et~al.(2013)Vijay, Paramanand, Rajagopalan, and
  Chellappa]{vijay2013non}
Channarayapatna~Shivaram Vijay, Chandramouli Paramanand, Ambasamudram~Narayanan
  Rajagopalan, and Rama Chellappa.
\newblock Non-uniform deblurring in hdr image reconstruction.
\newblock \emph{IEEE transactions on image processing}, 22\penalty0
  (10):\penalty0 3739--3750, 2013.

\bibitem[Purohit et~al.(2019)Purohit, Shah, and
  Rajagopalan]{purohit2019bringing}
Kuldeep Purohit, Anshul Shah, and AN~Rajagopalan.
\newblock Bringing alive blurred moments.
\newblock In \emph{Proceedings of the IEEE/CVF Conference on Computer Vision
  and Pattern Recognition}, pages 6830--6839, 2019.

\bibitem[Purohit and Rajagopalan(2020)]{purohit2020region}
Kuldeep Purohit and AN~Rajagopalan.
\newblock Region-adaptive dense network for efficient motion deblurring.
\newblock In \emph{Proceedings of the AAAI Conference on Artificial
  Intelligence}, volume~34, pages 11882--11889, 2020.

\bibitem[Mohan et~al.(2021)Mohan, Nithin, and Rajagopalan]{mohan2021deep}
MR~Mahesh Mohan, GK~Nithin, and AN~Rajagopalan.
\newblock Deep dynamic scene deblurring for unconstrained dual-lens cameras.
\newblock \emph{IEEE Transactions on Image Processing}, 30:\penalty0
  4479--4491, 2021.

\bibitem[Mohan et~al.(2019)Mohan, Girish, and
  Rajagopalan]{mohan2019unconstrained}
MR~Mohan, Sharath Girish, and AN~Rajagopalan.
\newblock Unconstrained motion deblurring for dual-lens cameras.
\newblock In \emph{Proceedings of the IEEE/CVF International Conference on
  Computer Vision}, pages 7870--7879, 2019.

\bibitem[Vasu et~al.(2018{\natexlab{a}})Vasu, Maligireddy, and
  Rajagopalan]{vasu2018non}
Subeesh Vasu, Venkatesh~Reddy Maligireddy, and AN~Rajagopalan.
\newblock Non-blind deblurring: Handling kernel uncertainty with cnns.
\newblock In \emph{Proceedings of the IEEE Conference on Computer Vision and
  Pattern Recognition}, pages 3272--3281, 2018{\natexlab{a}}.

\bibitem[Suresh and Rajagopalan(2007)]{suresh2007robust}
Kaggere~V Suresh and Ambasamudram~N Rajagopalan.
\newblock Robust and computationally efficient superresolution algorithm.
\newblock \emph{JOSA A}, 24\penalty0 (4):\penalty0 984--992, 2007.

\bibitem[Vasu et~al.(2018{\natexlab{b}})Vasu, Thekke~Madam, and
  Rajagopalan]{vasu2018analyzing}
Subeesh Vasu, Nimisha Thekke~Madam, and AN~Rajagopalan.
\newblock Analyzing perception-distortion tradeoff using enhanced perceptual
  super-resolution network.
\newblock In \emph{Proceedings of the European Conference on Computer Vision
  (ECCV) Workshops}, pages 0--0, 2018{\natexlab{b}}.

\bibitem[Bhavsar and Rajagopalan(2010)]{bhavsar2010resolution}
Arnav~V Bhavsar and AN~Rajagopalan.
\newblock Resolution enhancement in multi-image stereo.
\newblock \emph{IEEE transactions on pattern analysis and machine
  intelligence}, 32\penalty0 (9):\penalty0 1721--1728, 2010.

\bibitem[Bhavsar and Rajagopalan(2012)]{bhavsar2012range}
Arnav~V Bhavsar and Ambasamudram~N Rajagopalan.
\newblock Range map superresolution-inpainting, and reconstruction from sparse
  data.
\newblock \emph{Computer Vision and Image Understanding}, 116\penalty0
  (4):\penalty0 572--591, 2012.

\bibitem[Rajagopalan et~al.(2005)Rajagopalan, Chellappa, and
  Koterba]{rajagopalan2005background}
AN~Rajagopalan, Rama Chellappa, and Nathan~T Koterba.
\newblock Background learning for robust face recognition with pca in the
  presence of clutter.
\newblock \emph{IEEE Transactions on Image Processing}, 14\penalty0
  (6):\penalty0 832--843, 2005.

\bibitem[Rajagopalan and Kiran(2003)]{rajagopalan2003motion}
Ambasamudram~N Rajagopalan and V~Phani Kiran.
\newblock Motion-free superresolution and the role of relative blur.
\newblock \emph{JOSA A}, 20\penalty0 (11):\penalty0 2022--2032, 2003.

\bibitem[Nimisha and Rajagopalan(2021)]{nimisha2021blind}
TM~Nimisha and AN~Rajagopalan.
\newblock Blind super-resolution of faces for surveillance.
\newblock In \emph{Deep Learning-Based Face Analytics}, pages 119--136.
  Springer, 2021.

\bibitem[Purohit et~al.(2020)Purohit, Mandal, and
  Rajagopalan]{purohit2020mixed}
Kuldeep Purohit, Srimanta Mandal, and AN~Rajagopalan.
\newblock Mixed-dense connection networks for image and video super-resolution.
\newblock \emph{Neurocomputing}, 398:\penalty0 360--376, 2020.

\bibitem[Punnappurath et~al.(2017)Punnappurath, Nimisha, and
  Rajagopalan]{punnappurath2017multi}
Abhijith Punnappurath, Thekke~Madam Nimisha, and Ambasamudram~Narayanan
  Rajagopalan.
\newblock Multi-image blind super-resolution of 3d scenes.
\newblock \emph{IEEE Transactions on Image Processing}, 26\penalty0
  (11):\penalty0 5337--5352, 2017.

\bibitem[Punnappurath et~al.(2015)Punnappurath, Rengarajan, and
  Rajagopalan]{punnappurath2015rolling}
Abhijith Punnappurath, Vijay Rengarajan, and AN~Rajagopalan.
\newblock Rolling shutter super-resolution.
\newblock In \emph{Proceedings of the IEEE International Conference on Computer
  Vision}, pages 558--566, 2015.

\bibitem[Rengarajan et~al.(2017)Rengarajan, Balaji, and
  Rajagopalan]{rengarajan2017unrolling}
Vijay Rengarajan, Yogesh Balaji, and AN~Rajagopalan.
\newblock Unrolling the shutter: Cnn to correct motion distortions.
\newblock In \emph{Proceedings of the IEEE Conference on computer Vision and
  Pattern Recognition}, pages 2291--2299, 2017.

\bibitem[Rengarajan et~al.(2016{\natexlab{a}})Rengarajan, Rajagopalan, and
  Aravind]{rengarajan2016bows}
Vijay Rengarajan, Ambasamudram~N Rajagopalan, and Rangarajan Aravind.
\newblock From bows to arrows: Rolling shutter rectification of urban scenes.
\newblock In \emph{Proceedings of the IEEE Conference on Computer Vision and
  Pattern Recognition}, pages 2773--2781, 2016{\natexlab{a}}.

\bibitem[Kandula et~al.(2020)Kandula, Kumar, and Rajagopalan]{kandula2020deep}
Praveen Kandula, T~Lokesh Kumar, and AN~Rajagopalan.
\newblock Deep end-to-end rolling shutter rectification.
\newblock \emph{JOSA A}, 37\penalty0 (10):\penalty0 1574--1582, 2020.

\bibitem[Vasu et~al.(2018{\natexlab{c}})Vasu, Rajagopalan,
  et~al.]{vasu2018occlusion}
Subeesh Vasu, AN~Rajagopalan, et~al.
\newblock Occlusion-aware rolling shutter rectification of 3d scenes.
\newblock In \emph{Proceedings of the IEEE Conference on Computer Vision and
  Pattern Recognition}, pages 636--645, 2018{\natexlab{c}}.

\bibitem[Vasu et~al.(2017)Vasu, Rajagopalan, and Seetharaman]{vasu2017camera}
Subeesh Vasu, Ambasamudram~Narayanan Rajagopalan, and Guna Seetharaman.
\newblock Camera shutter-independent registration and rectification.
\newblock \emph{IEEE Transactions on Image Processing}, 27\penalty0
  (4):\penalty0 1901--1913, 2017.

\bibitem[Mohan et~al.(2017)Mohan, Rajagopalan, and Seetharaman]{mohan2017going}
Mahesh~MR Mohan, AN~Rajagopalan, and Gunasekaran Seetharaman.
\newblock Going unconstrained with rolling shutter deblurring.
\newblock In \emph{Proceedings of the IEEE International Conference on Computer
  Vision}, pages 4010--4018, 2017.

\bibitem[Rengarajan et~al.(2016{\natexlab{b}})Rengarajan, Rajagopalan, Aravind,
  and Seetharaman]{rengarajan2016image}
Vijay Rengarajan, Ambasamudram~Narayanan Rajagopalan, Rangarajan Aravind, and
  Guna Seetharaman.
\newblock Image registration and change detection under rolling shutter motion
  blur.
\newblock \emph{IEEE transactions on pattern analysis and machine
  intelligence}, 39\penalty0 (10):\penalty0 1959--1972, 2016{\natexlab{b}}.

\bibitem[Pichaikuppan et~al.(2014)Pichaikuppan, Narayanan, and
  Rangarajan]{pichaikuppan2014change}
Vijay Rengarajan~Angarai Pichaikuppan, Rajagopalan~Ambasamudram Narayanan, and
  Aravind Rangarajan.
\newblock Change detection in the presence of motion blur and rolling shutter
  effect.
\newblock In \emph{European Conference on Computer Vision}, pages 123--137.
  Springer, 2014.

\bibitem[Sun et~al.(2015)Sun, Cao, Xu, and Ponce]{sun2015learning}
Jian Sun, Wenfei Cao, Zongben Xu, and Jean Ponce.
\newblock Learning a convolutional neural network for non-uniform motion blur
  removal.
\newblock In \emph{Proceedings of the IEEE Conference on Computer Vision and
  Pattern Recognition}, pages 769--777, 2015.

\bibitem[Gong et~al.(2017)Gong, Yang, Liu, Zhang, Reid, Shen, Hengel, and
  Shi]{gong2017motion}
Dong Gong, Jie Yang, Lingqiao Liu, Yanning Zhang, Ian Reid, Chunhua Shen, AVD
  Hengel, and Qinfeng Shi.
\newblock From motion blur to motion flow: a deep learning solution for
  removing heterogeneous motion blur.
\newblock In \emph{The IEEE conference on computer vision and pattern
  recognition (CVPR)}, 2017.

\bibitem[Nah et~al.(2017)Nah, Kim, and Lee]{nah2017deep}
Seungjun Nah, Tae~Hyun Kim, and Kyoung~Mu Lee.
\newblock Deep multi-scale convolutional neural network for dynamic scene
  deblurring.
\newblock In \emph{CVPR}, volume~1, page~3, 2017.

\bibitem[Tao et~al.(2018)Tao, Gao, Shen, Wang, and Jia]{tao2018scale}
Xin Tao, Hongyun Gao, Xiaoyong Shen, Jue Wang, and Jiaya Jia.
\newblock Scale-recurrent network for deep image deblurring.
\newblock In \emph{Proceedings of the IEEE Conference on Computer Vision and
  Pattern Recognition}, pages 8174--8182, 2018.

\bibitem[Gao et~al.(2019)Gao, Tao, Shen, and Jia]{gao2019dynamic}
Hongyun Gao, Xin Tao, Xiaoyong Shen, and Jiaya Jia.
\newblock Dynamic scene deblurring with parameter selective sharing and nested
  skip connections.
\newblock In \emph{Proceedings of the IEEE Conference on Computer Vision and
  Pattern Recognition}, pages 3848--3856, 2019.

\bibitem[Zhang et~al.(2018)Zhang, Pan, Ren, Song, Bao, Lau, and
  Yang]{zhang2018dynamic}
Jiawei Zhang, Jinshan Pan, Jimmy Ren, Yibing Song, Linchao Bao, Rynson~WH Lau,
  and Ming-Hsuan Yang.
\newblock Dynamic scene deblurring using spatially variant recurrent neural
  networks.
\newblock In \emph{Proceedings of the IEEE Conference on Computer Vision and
  Pattern Recognition}, pages 2521--2529, 2018.

\bibitem[Ronneberger et~al.(2015)Ronneberger, Fischer, and
  Brox]{ronneberger2015u}
Olaf Ronneberger, Philipp Fischer, and Thomas Brox.
\newblock U-net: Convolutional networks for biomedical image segmentation.
\newblock In \emph{International Conference on Medical image computing and
  computer-assisted intervention}, pages 234--241. Springer, 2015.

\bibitem[Liu et~al.(2018{\natexlab{b}})Liu, Reda, Shih, Wang, Tao, and
  Catanzaro]{liu2018image}
Guilin Liu, Fitsum~A Reda, Kevin~J Shih, Ting-Chun Wang, Andrew Tao, and Bryan
  Catanzaro.
\newblock Image inpainting for irregular holes using partial convolutions.
\newblock In \emph{Proceedings of the European Conference on Computer Vision
  (ECCV)}, pages 85--100, 2018{\natexlab{b}}.

\bibitem[Isola et~al.(2017)Isola, Zhu, Zhou, and Efros]{isola2017image}
Phillip Isola, Jun-Yan Zhu, Tinghui Zhou, and Alexei~A Efros.
\newblock Image-to-image translation with conditional adversarial networks.
\newblock In \emph{Proceedings of the IEEE conference on computer vision and
  pattern recognition}, pages 1125--1134, 2017.

\bibitem[Yosinski et~al.(2014)Yosinski, Clune, Bengio, and
  Lipson]{yosinski2014transferable}
Jason Yosinski, Jeff Clune, Yoshua Bengio, and Hod Lipson.
\newblock How transferable are features in deep neural networks?
\newblock In \emph{Advances in neural information processing systems}, pages
  3320--3328, 2014.

\bibitem[Jang et~al.(2019)Jang, Lee, Hwang, and Shin]{jang2019learning}
Yunhun Jang, Hankook Lee, Sung~Ju Hwang, and Jinwoo Shin.
\newblock Learning what and where to transfer.
\newblock \emph{arXiv preprint arXiv:1905.05901}, 2019.

\bibitem[Hu et~al.(2018)Hu, Shen, and Sun]{hu2018squeeze}
Jie Hu, Li~Shen, and Gang Sun.
\newblock Squeeze-and-excitation networks.
\newblock In \emph{Proceedings of the IEEE conference on computer vision and
  pattern recognition}, pages 7132--7141, 2018.

\bibitem[Kupyn et~al.(2019)Kupyn, Martyniuk, Wu, and Wang]{kupyn2019deblurgan}
Orest Kupyn, Tetiana Martyniuk, Junru Wu, and Zhangyang Wang.
\newblock Deblurgan-v2: Deblurring (orders-of-magnitude) faster and better.
\newblock In \emph{Proceedings of the IEEE International Conference on Computer
  Vision}, pages 8878--8887, 2019.

\bibitem[Kupyn et~al.(2017)Kupyn, Budzan, Mykhailych, Mishkin, and
  Matas]{kupyn2017deblurgan}
Orest Kupyn, Volodymyr Budzan, Mykola Mykhailych, Dmytro Mishkin, and Jiri
  Matas.
\newblock Deblurgan: Blind motion deblurring using conditional adversarial
  networks.
\newblock \emph{arXiv preprint arXiv:1711.07064}, 2017.

\bibitem[Zhang and Patel(2018)]{zhang2018density}
He~Zhang and Vishal~M Patel.
\newblock Density-aware single image de-raining using a multi-stream dense
  network.
\newblock In \emph{Proceedings of the IEEE conference on computer vision and
  pattern recognition}, pages 695--704, 2018.

\bibitem[Li et~al.(2018)Li, Wu, Lin, Liu, and Zha]{li2018recurrent}
Xia Li, Jianlong Wu, Zhouchen Lin, Hong Liu, and Hongbin Zha.
\newblock Recurrent squeeze-and-excitation context aggregation net for single
  image deraining.
\newblock In \emph{Proceedings of the European Conference on Computer Vision
  (ECCV)}, pages 254--269, 2018.

\bibitem[Ren et~al.(2019)Ren, Zuo, Hu, Zhu, and Meng]{ren2019progressive}
Dongwei Ren, Wangmeng Zuo, Qinghua Hu, Pengfei Zhu, and Deyu Meng.
\newblock Progressive image deraining networks: a better and simpler baseline.
\newblock In \emph{Proceedings of the IEEE Conference on Computer Vision and
  Pattern Recognition}, pages 3937--3946, 2019.

\bibitem[Whyte et~al.(2012)Whyte, Sivic, Zisserman, and Ponce]{whyte2012non}
Oliver Whyte, Josef Sivic, Andrew Zisserman, and Jean Ponce.
\newblock Non-uniform deblurring for shaken images.
\newblock \emph{International journal of computer vision}, 98\penalty0
  (2):\penalty0 168--186, 2012.

\end{thebibliography}

\end{document}